\documentclass[preprint]{aastex}
\usepackage{graphicx}
\usepackage{threeparttable}
\usepackage{hyperref}
\usepackage{color}

\newcommand{\kms}{${\rm km \ s^{-1}}$}
\newcommand{\kmsb}{${\rm km \ s^{-1}}$ }
\newcommand{\ergs}{${\rm erg \ s^{-1}}$}
\newcommand{\ergsb}{${\rm erg \ s^{-1}}$ }
\newcommand{\msigma}{$\rm{M_{BH}}$--$\sigma_\star$ }

\begin{document}

\title{Keck/ESI Long-slit Spectroscopy of SBS 1421+511: A Recoiling Quasar Nucleus in An Active Galaxy Pair?}

\author{Luming Sun\altaffilmark{1,2}, Hongyan Zhou\altaffilmark{2,1}, Lei Hao\altaffilmark{3}, Peng Jiang\altaffilmark{1}, Jian Ge\altaffilmark{4}, Tuo Ji\altaffilmark{2}, Jingzhe Ma\altaffilmark{4}, Shaohua Zhang\altaffilmark{2}, Xinwen Shu\altaffilmark{5}}
\altaffiltext{1}{Department of Astronomy, University of Science and Technology of China, 96 Jinzhai Road, Hefei, Anhui, China, lmsun@mail.ustc.edu.cn}
\altaffiltext{2}{Polar Research Institute of China, 451 Jinqiao Road, Shanghai, China}
\altaffiltext{3}{Key Laboratory for Research in Galaxies and Cosmology, Shanghai Astronomical Observatory, Chinese Academy of Sciences, 80 Nandan Road, Shanghai 200030, China}
\altaffiltext{4}{Department of Astronomy, University of Florida, 211 Bryant Space Science Center, Gainesville, FL, USA}
\altaffiltext{5}{CEA Saclay, DSM/Irfu/Service dAstrophysique, Orme des Merisiers, 91191 Gif-sur-Yvette Cedex, France}

\begin{abstract}
  We present Keck/ESI long-slit spectroscopy of SBS 1421+511, a system consisting of a quasar at z = 0.276 and an extended source 3$\arcsec$ northern to the quasar.
  { The quasar shows a blue-skewed profile of Balmer broad emission lines, which can be well modeled as emissions from a circular disk with a blueshift velocity of $\sim$1400 \kms.
  The blueshift is better interpreted as resulting from a recoiling active black hole than from a super-massive black hole binary, since the line profile almost kept steady over one decade in the quasar rest-frame.
  Alternative interpretations are possible as well, such as emissions from a bipolar outflow or a circular disk with spiral emissivity perturbations.}
  The extended source shows Seyfert-like narrow line ratios and a [OIII] luminosity of $>1.4\times10^8L_\odot$, with almost the same redshift as the quasar and a projected distance of 12.5 kpc at the redshift.
  SBS 1421+511 is thus likely to be an interacting galaxy pair with dual AGN.
  Alternatively, the quasar companion only appears to be active but not necessarily so:
  the gas { before/in/behind} the companion galaxy is illuminated by the quasar as an extended emission line region is detected at a similar distance in the opposite direction southern to the quasar, which may be generated either by tidal interactions between the galaxy pair or large-scale outflows from the quasar.
\end{abstract}


\section{Introduction}

The merging of two galaxies will produce a binary black hole at the center of the newly formed galaxy.
Numerical relativity simulations predict that black holes can be kicked out of the galaxy in binary black hole mergers with velocities up to several thousand \kmsb \citep[e.g.][]{Campanelli07}.
Detection of a rapidly recoiling black hole would be a confirmation of the gravitational wave and shed new light on the cosmic evolution of super-massive black holes.
\cite{Komossa08} reported the first candidate, SDSS J092712.65+294344.0, which have two sets of emission lines with velocity difference of 2650 \kmsb and the red set has only narrow emission lines (NELs).
And since then several other recoiling black hole candidates have been reported \citep[e.g.][]{Shields09}.
However, all the candidates have interpretations other than recoiling black hole, such as peculiar double-peak emitters, close black hole binaries or chance superpositions \citep[e.g.][]{Bogdanovic09,Vivek09}.
There have been many efforts to search for recoiling black hole candidates in large area survey.
Most of them used shifted broad emission lines (BELs) relative to NELs as an indicator of recoiling black holes \citep[see][for a review]{Eracleous12}.
In addition, \cite{Volonteri08} suggested an ``off-nuclear AGN'' character, which means that a recoiling active black hole moves away from the center of its host galaxy, can also be used as the indicator of recoiling black holes.

We have run a systematic search for low-redshift rapidly recoiling black hole candidates based on SDSS data release 7.
The quasar catalog of \cite{Schneider07} was used as our parent sample.
We first selected those with $z<0.35$ so that complete H$\alpha$ profiles are available using SDSS data.
We then obtained the profiles of H$\alpha$ BELs by subtracting continuum and NELs.
After that, we detected the peaks from the smoothed profiles and measured their velocity shifts relative to NELs.
Quasars which show peaks on both blue and red sides are excluded because these are obvious double-peak emitters and difficult to be explained by recoiling black holes.
We found 22 sources with velocity shifts greater than 2000 \kms.
Among this sample, SBS 1421+511, which is a system consisting of a radio quasar at z = 0.276 and a nearby extended source (Figure 1(b)), is the only one showing an off-nuclear AGN character.
The position offset between the quasar and { the extended source} suggests that the quasar might be kicked out of its original host galaxy.
It can be seen from the SDSS spectrum (Figure 1(a)) that the profile of Balmer BELs is blue-skewed and the peak velocity of the profile is $\sim$3300 \kmsb relative to NELs.
Both the blue-skewed profile and the off-nuclear AGN character suggest a super-massive recoiling black hole.
If confirmed, SBS 1421+511 may shed new light on the study of recoiling black holes due to its off-nuclear AGN character which is rare among the candidates.

In this paper, we report the results of the new Keck-II Echellette Spectrograph and Imager (ESI) long-slit spectrographic observation of the system SBS 1421+511, which took the spectra of the quasar and the extended source at the same time.
We identified a host galaxy in the quasar position, { and therefore the off-nuclear AGN character is not due to a recoiling black hole.
The quasar BELs are well modeled as emissions from a circular disk with a blueshift velocity of $\sim$1400 \kmsb relative to NELs, suggesting a super-massive black hole which is recoiling at a velocity of $\sim$1400 \kms, and meanwhile other possible interpretations exist.
The physical justifications of these interpretations are discussed in Section 4.2.}

{ The spectrum of the extended source shows that it is an galaxy with a velocity difference relative to the quasar host galaxy of only 120 \kmsb and a projected distance of only 12.5 kpc at the redshift.
Therefore, the extended source is likely to be an companion galaxy and we refer to the extended source as ``the companion'' hereafter.}
The companion shows Seyfert-like line ratios and a high [OIII] luminosity, suggesting that SBS 1421+511 is a dual AGN system.
Dual AGNs have valuable potential as observational tools for studies of galaxy evolution, including measurements of galaxy merger rates, mass growth rates and merger rates of super-massive black holes \citep{Komossa03}.
[OIII] emission is often used as a indicator of AGN in surveys of dual AGNs.
For example, double-peak [OIII] NELs are used as signatures of dual AGNs in large area survey \citep[e.g.][]{Wang09} and separated [OIII] nuclei are important to confirming the dual AGN candidates \citep[e.g.][]{McGurk11}.
{ Intriguingly, we found an extended emission line region (EELR) southern to the quasar.
The existence of the EELR suggests that a similar being may coincide with the companion and the companion is not necessarily active.
We considered other possible interpretations such as a pair of EELRs with a bipolar structure and gas in the companion galaxy illuminated by the quasar, which are further discussed in section 4.3.}

{ The EELR itself is interesting.}
In spite of multiple observations, the nature and origin of EELRs \citep[e.g.][]{Stockton83} are still not understood completely and there is no generally interpretation for all the EELRs
\citep[see][for a review]{Fu09}.
They are initially thought to be tidal features, but later observations found that only a few of them show structures like tidal tails.
Only a small fraction of EELRs show accompanied large scale radio features, indicating that they are likely driven by radio jets \citep{Shih14}.
\cite{Fu09} presented that the luminosity of the EELR is strongly correlated with nuclear [OIII] luminosity, suggesting that most of the EELRs originate in AGN-driven galactic wind.
SBS 1421+511 is likely to be tidally interacting with a companion, and shows radio structures and large scale outflows at the same time, thus provide an ideal laboratory to study the nature of the EELR.
{ The possible origins of the EELR in SBS 1421+511 are discussed in Section 4.4.}

\section{Observation and Data Reduction}
We obtained the Keck ESI optical long slit spectrum of SBS 1421+511 on 2013 May 8 using the echelle mode.
The width of the slit is 0.75$\arcsec$ and the instrumental dispersion is $\sigma\sim$ 22 \kms.
The slit was aligned in a north-south direction (Figure 1(b)), so the quasar and the companion can fall into the slit at the same time.
The exposure time was 20 minutes.
After the exposure, we immediately obtained a spectrum of the standard star BD+33 2642 with the same setting and 2 minutes exposure to perform flux calibration and generate point spread function (PSF).
The seeing was $\sim1.4\arcsec$ measured from the Full Width of Half Maximum (FWHM) of the PSF.

We reduced the original data through a routine based on IDL program package \textbf{XIDL}\footnote{http://www2.keck.hawaii.edu/inst/esi/ESIRedux/index.html} to obtain two-dimensional (hereafter 2-D) spectral images.
Using the spectral images, we can easily obtain one-dimensional (hereafter 1-D) spectra by collapsing them along the spatial direction and obtain spatial brightness profiles along the slit by collapsing them along the spectral direction.
Bias and flat correction, wavelength calibration, sky subtraction were performed first.
Cosmic-rays were removed using \textbf{LACosmic} \citep{van Dokkum01}.
We resampled the clean spectral data around the tracks of the quasar for 10 echelle orders so that one pixel corresponds to 10 \kmsb in wavelength direction and 0.15$\arcsec$ in spatial direction, using 2-D wavelength calibration from Arc-lamp image and pixel scales from archival maps.
We then jointed the spectral images of the 10 orders.
The joint 2-D spectral image covers the wavelength range of 4000--10100 \AA\ after removing low response regions.
We then ran the same routine to obtain the spectral image of the standard star and then used the extracted 1-D spectrum of the standard star to make flux calibration of the source.
We also obtained the spatial brightness profiles in different wavelength ranges for the standard star to generate 1-D PSF.
There is a chance that the PSF may not be adequate for analysing the spectral image of the source due to seeing variance.
However, the variance is minor because the seeing was stable at that time according to the observation log and the time interval between the two exposures was only $\sim$12 minutes.

\section{Data Aanalysis and Results}

\subsection{The Nature of The Off-nuclear AGN Character}

SBS 1421+511 shows an off-nuclear AGN character.
It could be explained by a single-galaxy model that the quasar is naked and the companion is the original host galaxy of which the quasar might be kicked out.
There also exists an alternative double-galaxy model that the quasar has a host galaxy and the companion is indeed a companion galaxy.
{ Eracleous \& Halpern (2003, hereafter EH03) presented the analysis of optical spectra of SBS 1421+511 obtained with KPNO 2.1m and MDM 2.4m telescopes during 1997 and 2000.
According to the results of EH03, there is starlight emission underneath the quasar and the flux fraction of the starlight is 0.28$\pm$0.2.
The starlight can be interpreted as the host galaxy of the quasar according to the double-galaxy model, and as the outer part of the single galaxy according to the single-galaxy model.
We could not judge the two based on EH03's results because we had no spectrum of the companion and the starlight fraction has large uncertainty.}
We extracted the spectra of the quasar and the companion { from the new ESI 2-D spectral image} with 3$\arcsec$ apertures.
As can be seen from Figure 2, the quasar spectrum clearly shows absorption features of old stellar populations such as CaII K band around 3934 \AA, Mg Ib feature around 5177 \AA, Na ID $\lambda\lambda$ 5890, 5896 doublet lines, { proving that} there is starlight underneath the quasar.
And the companion spectrum shows strong these features, and thus is dominated by old stellar populations.
The absorption line redshifts are close for the two spectra.
{ According to the single-galaxy model, the stellar emission in the quasar spectrum comes from the outer part of the single galaxy and thus its flux would be much lower than that in the companion spectrum, which is extracted using the same aperture size.
And it is not necessarily so according to the double-galaxy model.}
Therefore a decomposition of quasar and stellar components in the quasar spectrum may help to distinguish the two models.

\subsubsection{Spectral Decomposition}

We fit the quasar spectrum using models consisting of a stellar component and a quasar component.
The stellar component is the addition of simple stellar population (SSP) templates of \cite{Bruzual03}, which were broadened by convolving with a Gaussian to match the stellar velocity dispersion and shifted to match the redshift.
We used a single SSP and the best-fitting result was achieved with stellar population of $\sim$3 Gyr old, and redshift and velocity dispersion are $z=0.27626\pm0.00007$ and $\sigma_\star = 85\pm20$, respectively.
We attempted to add one more SSP but after this the reduced $\chi^2$ only varied 0.02 and the second SSP contributes little flux, and therefore we did not adopt the second SSP.
For the quasar component, we attempted to use a single power-law as the quasar continuum at first and the residual showed excess in spectral region redder than 6000 \AA.
The excess is also seen in quasar composite spectra \citep{Vanden Berk01} and is interpreted as the radiation of a torus.
Therefore we used a broken power-law with broken point fitted to be $\sim$5700 \AA\ for the quasar continuum.
We masked the spectral regions affected by NELs, which are demonstrated in Figure 2.
There are strong and wide BELs, including Balmer series and HeI $\lambda$5876, which influence some important stellar features, and therefore we fit the continuum and BELs at the same time instead of masking the regions affected by BELs in the fitting.
The BEL models are described as follows:
1. We used double Gaussian for all the BELs since they are obviously asymmetric.
2. The residual reveals a feature at around 4700 \AA.
Though the quasar composite spectra show FeII pseudo continuum here, it is unlikely that the feature is due to FeII emission because the data were not well fitted by accordingly adding an optical FeII template.
Thus we considered another possibility that the feature is HeII $\lambda$4686 BEL, which is seen in the spectra of FeII weak quasars.
We added a corresponding double Gaussian for HeII of which the profile is assumed to be identical to that of H$\beta$, and yielded a good fit to the data.
3. The residual shows an excess at $<4000$ \AA, which is likely due to blended high order Balmer BELs and thus we added a component accordingly.
The component consists of Balmer lines from $n=7$ to $n=50$, of which the profiles are fixed to that of H$\delta$, and of which the relative strength are fixed using the line emissivity for Case B \citep{Storey95} under a typical circumstance for broad line regions of $T_e=15,000$ K and $n_e=10^9$ cm$^{-3}$.

We also used SSP templates to fit the companion spectrum.
We tried single SSP and double SSP, and the results showed that double SSP is appropriate and the ages of the two SSPs are $\sim2$ Gyr and $\sim$10 Myr.
The redshift and velocity dispersion of the SSPs are $z=0.27576\pm0.00007$ and $\sigma_\star = 95\pm20$ \kms, respectively. 
Considering that the companion spectrum is contaminated by the quasar continuum and BELs due to seeing effect, we added a contamination component which is $\sim$4\% of the corresponding flux in the quasar spectrum and the fraction was computed using PSF.

We overplot the best-fitting models for the two spectra in Figure 2.
{ The fraction of starlight increases along direction of longer wavelength from 19\% at 4000 \AA\ to 46\% at 7000 \AA, in agreement with EH03's result.}
The { corresponding stellar mass} in the quasar spectrum is $5\times10^{10}M_\odot$, close to that in the companion spectrum of $4\times10^{10}M_\odot$.
Therefore it is unlikely that the starlight underneath the quasar comes from the outer part of the single galaxy.
The stellar mass is high thus it is also unlikely that the stellar component comes from stars which are ejected accompanied with the super-massive black hole when recoiling.

\subsubsection{Spatial Decomposition}

We display the 1-D spatial brightness profile of the continuum from Keck data in Figure 3(a).
The profile was extracted from a spectral region at around rest-frame 7000 \AA, where the contribution of the quasar component is relatively small.
As illustrated by Figure 3(a), we obtained the minimum flux of the non-quasar component (orange line). 
We also show the 2-D brightness profile from SDSS i-band image in Figure 3(c).
This band (rest-frame 5400--6500 \AA) was selected because it is not strongly affected by emission lines.
Using PSF from a nearby star, we obtained a similar minimum flux (Figure 3(d)) and the inferred lower limit of the stellar luminosity is $V<-22.7$\footnote{We have corrected for spectral shape.}.
Such a bright galaxy would be either an elliptical or a disk galaxy if not a merger, and the radial brightness profiles of the both can be well described by S\'{e}rsic profile.
Thus this profile is used to fit the continuum brightness profile for both the 1-D (Keck) and 2-D (SDSS) data, and the later was done with the code \textbf{GALFIT} \citep[version 3.0,][]{Peng10}.
We used two models: a single-galaxy model consisting of a S\'{e}rsic profile for the single galaxy and a point source for the quasar component, and a double-galaxy model with one more S\'{e}rsic profile for the host.
For 1-D data the double-galaxy model is { significantly better} than the single-galaxy model since the $\chi^2$/dof reduces from 65603/115 to 2505.2/110, { and for 2-D data it is marginally better} since the $\chi^2$/dof are 11092.8/12304 and 11060.8/12297.
{ We adopted the conclusion from 1-D data that} the single-galaxy model was excluded { as the total signal to noise ratio of 1-D data is much higher than that of 2-D data.}
From 1-D spatial decomposition results, we obtained the contribution of the stellar component { in an aperture which was used to extract quasar spectrum at around 7000 \AA\ }to be 44\%, in accordance with that obtained from spectral decomposition results of 46\%.
The good agreement makes the two decompositions more robust.
{ Therefore we conclude that the off-nuclear AGN character is not due to a recoiling super-massive black hole but due to a close companion.}


The distance between the centers of the two galaxies is $d=3.1$ arcsec\footnote{
We do not list the errors for the 1-D data because the statistical error is too small and the error is dominated by model uncertainty.}
and $d=3.0\pm0.1$ arcsec for 1-D and 2-D data, respectively, which are consistent.
{ The little distance} (12.5 kpc in projection at $z=0.276$) and { the little radial velocity difference} (120 \kms) indicate that the two galaxies are likely physically connected and rotating around each other, and going on a major merger considering that both the two galaxies are bright with $V_{\rm host}=-22.3\pm0.3$ and $V_{\rm companion}=-22.5\pm0.1$\footnote{
These values have been corrected for spectral shape using the best-fitting stellar components for the quasar and companion spectra.}
, which were obtained from 2-D spatial decomposition results.
The S\'{e}rsic index of the companion galaxy is $n=1.1$ and $n=1.3\pm0.2$ for the 1-D and 2-D fit, respectively, being close to 1 and implying that the companion galaxy may be a disk galaxy.
On the other hand, the S\'{e}rsic index of the host galaxy is $n=1.9$ for 1-D fit, and the corresponding value for 2-D fit is undetermined due to large uncertainty, { and we are not sure if it is a disk galaxy or an elliptical galaxy.}

\subsection{Analyzing the Blue-skewed profile of BELs}
\label{3.2}

The blue-skewed profile of quasar BELs of SBS 1421+511 implies a super-massive recoiling black hole.
However, similar profiles are also seen in some AGNs with double-peaked BELs.
{ EH03 analyzed the H$\alpha$ BEL profile of SBS 1421+511 using spectrum obtained with MDM 2.4m telescopes in 1998.
They fit the line with the relativistic Keplerian disk model of \cite{Chen89}, which is widely used for double-peaked BELs.
The model assumes that the line originates on the surface of a circular, Keplerian disk whose axis is inclined by an angle $i$ relative to the line of sight, between radii $\xi_1$ and $\xi_2$ (expressed in units of the gravitational radius, $r_g\equiv G M_{\rm BH}/c^2$, where $M_{\rm BH}$ is the mass of the black hole).
The disk has an axisymmetric emissivity of the form $\epsilon \propto \xi^q$.
Local broadening of the line is represented by a Gaussian rest-frame profile of velocity dispersion $\sigma$.
The free parameters of the model are thus $\xi_1$, $\xi_2$, $i$, $q$ and $\sigma$.
EH03 found that the H$\alpha$ BEL could not be well fitted with a simple circular disk model, and the fit could be improved by either (1) applying an ad hoc blueshift to the model of 1370 km/s or (2) using an elliptical disk model, which has two more free parameters of eccentricity $e$ and the angle between the major axis of the elliptical disk and the line of sight $\phi_0$ \citep{Eracleous95}.
We accordingly fit the H$\alpha$ profile of ESI, which was obtained by subtracting continuum from original quasar spectrum, using both circular and elliptical disk models.
We also attempted to apply a blueshift to the models.
The statistics and best-fitting parameters of the models are listed in Table 1.
Without blueshift, neither circular nor elliptical disk model can well describe the H$\alpha$ profile.
After applying an ad hoc blueshift, both the two models yield good fit, and they are statistically just as good, for $\chi^2$/dof is 1013/843 and 1004/841 for the two models.
The best-fitting circular disk model with a blueshift is displayed in the upper panel of Figure 4.
The ad hoc blueshift applied to the circular disk model is 1450 \kms, close to the value obtained by EH03.
We confirmed EH03's conclusions that the double-peaked BELs of SBS 1421+511 can not be well reproduced by a simple circular disk model and the fit was greatly improved by applying an ad hoc blueshift.
And we excluded the elliptical disk model because: 1. it can not reproduce the profile without a blueshift; 2. it is statistically not better than circular disk model if applying a blueshift.

Long term monitoring of AGNs with double-peaked BELs \citep{Gezari07,Lewis10} reveal profile variations, which can be described by the changes of excess emissions in amplitude and projected velocity on timescales of years.
Since we have three spectra of SBS 1421+511 from MDM (1998), SDSS (2003) and Keck (2013), we plotted the normalized H$\alpha$ profiles of the three spectra together in the lower panel of Figure 4 to see the variability of the profile.
The variance between the SDSS and Keck profiles is little, and the blue peak in MDM profile is weaker than those in SDSS and Keck profiles.
We did not find velocity change of the blue peak among the three observations, though the change is common among AGNs with similar BEL profiles.
To quantify this, we estimated the change by comparing MDM and SDSS profiles with shifted and normalized Keck profile and by using minimum Chi-square method.
The change is $-20\pm60$ \kmsb between SDSS and Keck profiles, and $-60\pm60$ \kmsb between MDM and Keck profiles, and the inferred velocity change rate is $dv/dt=-2\pm4$ \kmsb per year in rest frame.
This constitutes an important constraint on the nature of the double-peaked BEL of SBS 1421+511, and we discuss the possible interpretations of the BELs in Section 4.2.}

\subsection{Extended Emission Line Features}
\label{3.3}

It can be seen from Figure 2 that the companion spectrum shows detectable [HeII] $\lambda$4686 and strong [OIII] $\lambda\lambda$4959, 5007.
The line ratio of [OIII]/H$\beta$ is $\sim7$ measured from the residual between the original spectrum and the best-fitting model.
These suggest that the gas in the companion is ionized by an accreting black hole and the companion may also be an AGN.
The [OIII] emission of the companion is so strong that a corresponding emission line feature can be easily seen northern to the quasar from the 2-D emission line spectral image (Figure 5(a)).
Intriguingly, there also exists an emission line feature in the opposite direction southern to the quasar.
{ As can be seen from the 2-D spectral image, the two features show symmetry in both spatial and velocity directions: the northern one is blueshifted relative to the quasar host galaxy and the southern one is redshifted, and the distances to the quasar position are both several arcsecs and the velocity differences are both $v\sim200$ \kms.}
As the symmetry we investigated the two features together.
For clarity, we referred to the northern one as ``Feature N'' and the southern one as ``Feature S'' hereafter.

We extracted the spatial brightness profiles of Feature N using a velocity range of $-300<v<-100$ \kms, which are presented in Figure 5(b).
Both the peak positions of profiles for [OIII] and [NII] locate $\sim$3$\arcsec$ north to the quasar, roughly coincident with the center of the companion (horizontal yellow dashed line), and for H$\alpha$ and H$\beta$ the peak positions are $\sim$4\arcsec\ north to the quasar.
{ The disparity} indicates that Feature N may consist of two components: one has stronger [OIII] and [NII] emissions whose major emitting region is around the center of the companion; the other has stronger Balmer emissions whose major emitting region is in the outer region, and hence likely comes from an off-center HII region.
Considering the 1.4$\arcsec$ seeing, the two components are mixed with each other.
{ To decompose them} we extracted 1-D spectra (Figure 5(d)) using two 2$\arcsec$-width apertures, AperN (blue in Figure 5(c)) and AperHII (purple).
{ We selected AperN whose center is 3$\arcsec$ north to the quasar, in consistent with the major emitting region of the first component, and selected AperHII so that the emission most comes from the second component.
We obtained the emission line spectra by subtracted the continua\footnote{
The continua were obtained by fitting spectra in non-emission-line regions using SSPs and quasar contamination components as what we did for the companion spectrum in Section 3.1.1.},
and the velocity profiles of [OIII] and H$\alpha$ are plotted in Figure 5(e).}
We also classified the type of ionization source using the flux ratios of emission lines, which are listed in Table 2 and plotted in Figure 5(f), according to the diagnostic diagram of \cite{Kewley06}.
For AperN, the line ratios obtained by simply performing a Gaussian fitting are Seyfert-like (blue star in Figure 5(f)) according to the diagnostic diagram.
Considering the contamination of the second component, which contributes more to H$\alpha$ and H$\beta$ than to [OIII] and [NII], the line ratios of the first component would move upper right on the diagnostic diagram and still be AGN-like.
For AperHII, [OIII] profile is nearly the same as that of AperN (Figure 5(e)), suggesting that [OIII] emission is dominated by the first component.
By contrast, H$\alpha$ profile is much narrower than that of AperN, and for both the two apertures H$\alpha$ profiles are narrower than [OIII] profiles, implying that H$\alpha$ emission of AperHII is dominated by the second component and H$\alpha$ emission of AperN is the mixture of the two components.
By directly fitting the profiles using Gaussian, the line ratios of AperHII (purple star) are close to the dividing line of AGN and HII regimes in the diagnostic diagram.
Furthermore, the line ratios (purple diamond) are HII-like if calculated using fluxes which are integrated over a narrower velocity range of $-240<v<-80$ \kms, { which is less contaminated}.
Thus it is likely that the second component has HII-like line ratios.
We then decomposed the emission lines of the two apertures into two sets which are corresponding to the two components by assuming that the profiles { are Gaussian} and are identical for all the emission lines for each set.
The results show that 87\% and 13\% of the first set fall into AperN and AperHII, respectively, and the corresponding fractions of the second set are 44\% and 56\%.
The line ratios of the first set (blue square) are Seyfert-like, and those of the second set (purple square) are HII-like.
The velocity difference of the Seyfert-like set of emission lines relative to the absorption line of the companion galaxy is $-70$ \kmsb and that of the HII-like set is $-40$ \kmsb, and the velocity dispersions are 75 and 150 \kmsb for two sets, respectively.
Therefore we concluded that Feature N is composed of { a component in the companion center which has Seyfert-like line ratios} and an off-center HII region.
{ The [OIII] luminosity of the Seyfert-like component in a rectangular aperture which is set by 0.75$\arcsec$ slit and 2$\arcsec$ AperN} is $L=6\times10^7\ L_\odot$.
The contamination by quasar emission due to seeing effect is minor as the fraction is $\sim10\%$ predicted by PSF.
{ To obtain the total luminosity of the source of the Seyfert-like component, we should multiply} an aperture correction which is computed to be 2.3 by using PSF and assuming the source is point-like, and higher if the source is extended.
Thus the total [OIII] luminosity is L$_{\rm{[OIII]}}> 1.4\times10^8L_\odot$, which may be underestimated considering an uncertain extinction correction.
Such a high [OIII] luminosity suggests an powerful AGN in the companion galaxy.

We also extracted the spatial brightness profiles of feature S using a velocity range of $100<v<300$ \kms. Feature S shows high [OIII] flux relative to H$\beta$ and extends to at least 6$\arcsec$ (25 kpc in projection) away from the quasar where little starlight is seen, indicating an EELR.
In order to compare the properties of Feature S and Feature N, we extract a 1-D spectrum in an aperture AperS (red in Figure 5(c,d)) whose width and distance to the quasar are both the same with those of AperN.
The velocity profile of [OIII] shows a broad base, which may come from the contaminations by emissions of the quasar.
To mask out the base, we fit [OIII] emission line using double Gaussian and { subtract the blueshifted broad Gaussian component.
By assuming that the rest emission comes from an EELR, we obtained the velocity difference and the velocity dispersion of the EELR of $v=170$ \kmsb and $\sigma=160$ \kms.
The line ratios of AperS spectrum (red star in Figure 5(f)) are Seyfert-like, agreeing with that the emission comes from an EELR.}

We will further discuss the natures of Feature N and Feature S in Section 4.3 and 4.4.
For the discussion, we estimated the electron density of the two features using line ratios of [SII] $\lambda$6716/$\lambda$6731 and [OII] $\lambda$3729/$\lambda$3726, and electron temperature using line ratio of [OIII] ($\lambda$4959+$\lambda$5007)/$\lambda$4363 \citep{Osterbrock06}.
For feature N, we obtained only an upper limit of the electron density of $n_e<100$ cm$^{-3}$ (95.3\% confidence) according to [SII] ratio of $1.39\pm0.05$ and [OII] ratio of $1.52\pm0.07$, and the electron temperature is $T_e\sim20,000$ K according to [OIII] ratio of $39\pm4$.
For feature S, the electron density of Feature S is $n_e=120\pm60$ cm$^{-3}$ according to the [SII] ratio of $1.30\pm0.08$, and the uncertainty of [OII] line ratio is too large to limit the density.
[OIII] $\lambda$4363 line is not detected in AperS spectrum, and the inferred electron temperature is $t_e<20000$ K (95.3\% confidence) using the upper limit.
The electron density is higher in AperS than in AperN and the electron temperature is lower.

\subsection{the Blue Wing in the NEL profiles of the Quasar Spectrum}
\label{3.4}

In the quasar spectrum the profiles of NELs show strong blue wings, which are generally thought to be correlated with an emission line outflow.
Figure 6(b) displays the velocity profiles of three isolated NELs (H$\beta$, [OIII] $\lambda$5007 and [OI] $\lambda$6300).
Besides relatively narrow core components, strong blue wing components are seen in both high- and low-ionized lines, which extend for a velocity up to 1200 \kmsb measured from the [OIII] velocity profile, suggesting a strong outflow.
We tried to parameterize the NELs to further investigate the two components.
We used double Gaussian for all the NELs: one for the core component and the other for the blue wing component.
Some lines are blended, such as H$\alpha$+[NII] $\lambda\lambda$6548,6583 lines and [SII] $\lambda\lambda$6716,6731 doublet, so we tied the parameters of these lines as follows to reduce the uncertainty.
For [NII] doublet, the profiles were assumed to be identical and the flux ratios were fixed to theoretical value 2.96.
And for [SII] doublet, the centroids were tied for both the core component and the wing component, and the widths were also tied, while the doublet flux ratios were set to be free parameters.
The two Gaussians to fit [OIII] are shown in Figure 6(b) (purple lines).
The redshift and the velocity dispersion of the core component are $z=0.27620$ and $\sigma=90$ \kms, respectively, consistent with those of the starlight for the quasar spectrum, indicating that the kinematics of the core component is dominated by bulge gravitation and this component is likely to represent a traditional narrow line region (NLR).
The wing component contributes 70\% of the total flux for [OIII], and the velocity difference with the core component is $-200$ \kms, and the velocity dispersion is 340 \kms.
Compared with the core component, the wing component is more complex, especially the near-zero-velocity part of this Gaussian, which can be understood in two ways.
One is that it comes from the traditional NLR { whose velocity distribution} is more extended than a Gaussian profile.
The other is that it does not come from the traditional NLR, though it has a radial velocity close to zero for some reason.
In order to avoid confusion, we used a velocity range of $-1200<v<-400$ \kms where the emission mostly comes from the outflow according to both interpretations, instead of the blueshifted Gaussian, when referring to the blue wing component.
We accordingly used an velocity range of $-200<v<200$ \kmsb when mentioning the core component.

We noticed that the blue wing component shows lower ionization state than the core component as the line ratios of [OIII]/H$\beta$ are $7.1\pm0.1$ and $9.9\pm0.1$ for the two component.
The situation is different from blue wing outflows in most of other AGNs, which show blue wings stronger in high-ionized lines than in low-ionized lines.
{ Since the line ratios of [OIII]/H$\beta$ are high, it is likely that both the outflow material and traditional NLR in SBS 1421+511 are mainly ionized by the quasar photons.
If so, considering that the electron densities from [SII] are close for the two component, the lower ionization state suggests that the outflow region in SBS 1421+511 may be more extended than traditional NLR, which typically has a size of sub kpc to several kpc.
This implies that the outflow region may have a size at an magnitude of order of $\sim$kpc, and thus may be resolvable under 1.4 arcsec seeing condition.}
To check this we extracted the spatial brightness profile of the blue wing component (Figure 6(c)).
Although the northern half is influenced by Feature N (see Figure 6(a)), we can see from the uncontaminated southern half that the profile is broader than the PSF.
This is impossible to be caused by seeing variance because the profile is also broader than that of the core component.
Therefore we conclude that the outflow region is resolved.
Considering the seeing FWHM corresponds to $\sim$6 kpc, the distinct disparity between the spatial profile and that of a point source indicates that the size of the outflow region is comparable to the seeing value, and thus is possibly at an order of magnitude of kpc.
We then estimated the size of the blue wing outflow by fitting the southern half of the spatial profile.
We tried several profile models, such as double Gaussian, exponential and S\'{e}rsic, and found that a S\'{e}rsic profile model (green lines in Figure 6(d)) with index $n=1.96$ fit the profile best.
The size of the outflow region is then estimated to be $\sim$1 kpc using the half-light radius of the model profile.


We estimated the mass rate of the blue wing outflow with a method used to estimate the parameters of NLR \citep{Peterson97}.
The mass of emitting gas can be calculated as
$M\sim7\times10^5\frac{L_{41}(\rm H\beta)}{n_{e,3}}M_{\odot}$,
where $L_{41}(\rm H\beta)$ is the H$\beta$ luminosity in unit of $10^{41}$ \ergs, and $n_{e,3}$ is the electron density in unit of $10^3$ cm$^{-3}$.
The mass outflow rate is then $\dot{M}\sim Mv/R$, where $v$ is the mean velocity of the outflow material and $R$ is the size of outflow region.
Using the profile in the velocity range of $-1200<v<-400$ \kms, the H$\beta$ luminosity\footnote{
We have made extinction correction using Balmer decrement and aperture correction using PSF.}
is L(H$\beta$)$=4\times10^{41}$ \ergsb and the mean velocity is $v=660$ \kms.
The size of outflow region is estimated to be 1 kpc from the spatial profile and the electron density is $2.3\pm0.2$ from the flux ratio of [SII] $\lambda$6716/$\lambda$6731 of the wing component of $1.21\pm0.06$.
Using these values, the mass of the emitting material is $M=2\times10^7\ M_{\odot}$ and the inferred mass outflow rate is $\dot{M}\sim13 M_{\odot}\ \rm{yr}^{-1}$.
The kinetic energy outflow rate is $\dot{E}=\frac{1}{2}\dot{M}v^2\sim2\times10^{42}$ \ergs.

\section{Discussion}

\subsection{The Bolometric luminosity and The Black Hole Mass of SBS 1421+511}
\label{4.1}

{ The bolometric luminosity and the black hole mass are key parameters to understand the quasar in SBS 1421+511.}
We estimated the bolometric luminosity by applying bolometric correction to both UV-optical continuum luminosity and hard X-ray luminosity.
The luminosity indicators of L$_{2500}$ and L$_{2{\rm keV}}$\footnote{
monochromatic luminosity in rest-frame 2500\AA\ and 2 keV} were used since { we could obtain the ratio of optical and X-ray $\alpha_{\rm OX}$ to compare with other AGNs.}
Because neither the SDSS nor the Keck spectrum covers a rest-frame wavelength of 2500\AA\footnote{
The blue ends of the wavelength coverage are 2980\AA\ and 3130\AA\ for the two spectra.}, we obtained L$_{2500}$ by extrapolating the continua in spectral regions bluer than 3500\AA.
The aperture corrected L$_{2500}$ for both the two spectra are $\sim3\times10^{44}$ \ergs, and the variance between them is less than 10\%.
SBS 1421+511 was observed by Swift X-Ray Telescope (XRT) in 2006 April and 2010 July (Target ID 35375),
and the effective exposure time was 5.2 and 2.7 ks for the two observations, respectively.
We extract the spectra for PC mode using \textbf{Xselect} (Version 2.4) through standard routine and combined the two spectra.
We then grouped the combined spectrum so that it has at least 20 counts per bin to ensure the $\chi^2$ statistics, and fit it using \textbf{XSpec} (Version 12.8).
The spectrum is well fitted ($\chi^2=24.6/30$) by an absorbed power law with index $\Gamma=1.85\pm0.10$ and neutral absorption column density $N_{\rm H}<4\times10^{20}$ cm$^{-2}$ (90\% confidence).
The L$_{2{\rm keV}}$ is $2.5\times10^{44}$ \ergsb and the inferred $\alpha_{\rm OX}$ is 1.1 for SBS 1421+511, lower than 1.33, the typical value of radio loud quasars \citep{Elvis94}.
We checked whether this is due to variability in optical or in X-ray.
The variance of L$_{2500}$ between 2003 (SDSS) and 2013 (Keck) is $<$10\%, and the variance of L$_{2{\rm keV}}$ between 2006 and 2010 is $\sim$30\%, which was obtained by fitting the data of two Swift observations individually.
Assuming these are the variability amplitudes of the quasar, the uncertainty of $\alpha_{\rm OX}$ due to variability is only 0.06.
Therefore the unusual shape of spectral energy distribution (SED) is barely attributed to variability.
On the other hand, SBS 1421+511 is a flat spectrum radio quasar (FSRQ) with spectral index $\alpha=0.4$ (defined as $F_\nu\propto\nu^{-\alpha}$) and P$_{1.4{\rm GHz}}=5\times10^{25}$ W Hz$^{-1}$, which is resolved in Very Large Baseline Array (VLBA) observation.
We show in Figure 1(c) the brightness distributions map\footnote{
Obtained from \textit{Astrogeo Center Database of Brightness Distributions, Correlated Flux Densities, and Images of Compact Radio Sources Produced with VLBI} (http://astrogeo.org/vlbi\_images/). We thank G. J. Taylor for kindly permitting us to use this image.}
taken during VLBA Imaging and Polarimetry Survey \citep[VIPS][]{Helmboldt07}.
The analysis of Helmboldt et al. demonstrated that the radio source has a jet structure { with a size of $>$30 pc} and a position angle of $-33.7$ degrees (north by east).
The power of the jet can be estimated with the radio luminosity \citep{Merloni07,Birzan08} to be $2\times10^{44}$ \ergs, similar to the X-ray luminosity, and thus the excess in X-ray luminosity { relative to UV-optical} may be due to radio jet activity.
Accordingly, we used quasar 3C206 from the sample of Elvis et al. whose $\alpha_{\rm OX}$ (1.15) is the closest to that of SBS 1421+511 as SED template, and estimated the bolometric luminosity to be $3\times10^{45}$ \ergs.

{ The central black hole mass of the quasar in SBS 1421+511 could not be estimated from the BELs because they are double-peaked.}
Using the stellar velocity dispersion of the host galaxy ($\sigma_\star=85\pm20$ \kms) and \msigma relation \citep{Greene06}, we estimated the black hole mass to be $4\times10^6M_\odot$, and the inferred Eddington ratio is 5.
{ The measurement of $\sigma_\star$ is robust considering that it is consistent with the velocity dispersion of gas in the bulge, which is measured to be 90 \kmsb from the core components of NELs.
The uncertainty of black hole mass is dominated by the intrinsic scatter of the \msigma relation, which is 0.5 dex for 1$\sigma$ confidence.
We also estimated the black hole mass to be $2\times10^7M_\odot$ using the bolometric luminosity and assuming that the black hole is accreting at Eddington accretion rate.}

The kinetic energy outflow rate of the kpc-scale outflow is $\sim0.06\%$ of the AGN luminosity in SBS 1421+511, and hence is rather strong.
{ The main source driving the outflow can not be starburst because the maximum velocity of the outflow $\sim1200$ \kmsb is higher than those driven by starbursts \citep{Rupke05}.
It is also unlikely to be the radio jet because the scale of radio jet is one order of magnitude smaller than the size of outflow.}
Ram pressure from AGN wind and radiation pressure on the dust are also suggested as the possible origins of the AGN driven outflow, and these mechanisms are effective when the AGN luminosity is close to Eddington luminosity \citep{Fabian12}.
Thus { it is likely that} AGN wind or radiation pressure drives the kpc-scale outflow in SBS 1421+511.

\subsection{The Nature of the Blue-skewed BEL profiles}
\label{4.2}

{
In section 3.2, we attempted to reproduce the blue-skewed H$\alpha$ profiles of SBS 1421+511 with accretion disk models, which are the most successful models for double-peaked BELs till now.
A simple circular disk model of \cite{Chen89} is unable to interpret the data.
This can be easily understood, since a circular disk produces blue and red peaks with similar fluxes while in SBS 1421+511 only the blue peak is seen.
Thus there must be an excess emission to the circular disk which is blueshifted in velocity and contributes to the strong blue peak, and hence the disk is nonaxisymmetric.
There are multiple nonaxisymmetric disk models such as elliptical disk \citep{Eracleous95}, circular disk with an orbiting bright spot \citep[e.g.][]{Newman97} and circular disk with spiral emissivity perturbations \citep{Storchi-Bergmann03}. 
The elliptical disk model is first ruled out because it can not reproduce the data.
Although we did not directly examine other nonaxisymmetric models by fitting the data, we can test them using variability behavior of the BEL profile because the excess emission from a nonaxisymmetric structure leads to long-term periodic profile variability.
For the bright spot model the period is the dynamic time scale, which can be estimated using the peak velocity of H$\alpha$ profile ($\sim$3000 \kms) and the black hole mass.
The period is $<$4 yrs and $<$20 yrs for black hole masses of $4\times10^6M_\odot$ and $2\times10^7M_\odot$, respectively, and 1.5 and 7 yrs assuming the unknown inclination angle to be 45$^\circ$.
For spiral model the period is the precession time scale, which is limited by sound-crossing time scale and is typically several times higher than the dynamic time scale \citep{Storchi-Bergmann03}, and thus is at an order of magnitude of decades for SBS 1421+511.
These period values agree with those obtained from monitoring of AGNs with double-peaked BELs, which are months to decades \citep{Gezari07,Lewis10}.
On the other hand, the period $P$ can be limited observationally using the radial velocity of the excess emission $v_r$ and the velocity change rate $\rm{d}\it{v}_r/\rm{d}\it{t}$:

$P = \frac{v_r \cdot \rm{sin}\it{i} \cdot \rm{sin}\it{\phi}}{\rm{d}\it{v}_r/\rm{d}\it{t}}$,

where i is the inclination angle of the disk and $\phi$ is the phase angle in the disk of the source which produces the excess emission.
Using the peak velocity of H$\alpha$ profile as $v_r$ and 3$\sigma$ upper limit of $\rm{d}\it{v}_r/\rm{d}\it{t}$ and assuming the unknown angles to be 45$^\circ$, the observed lower limit of the period is $\sim$100 yrs.
Thus the bright spot model is unlikely and the possibility of spiral model remains open considering the uncertainty in black hole mass, unknown angles and structure of accretion disk.

We showed that the data can be well reproduced using circular disk model if applying an ad hoc blueshift with a velocity of 1400 /kms relative to NELs.
This is unusual since the redshifts of disk models are typically fixed at those of host galaxies in previous research works.
The large blueshift velocity can be physically interpreted either as resulting from orbiting around an unseen super-massive black hole companion or as resulting from recoiling at a large velocity.
The orbiting model predicts that the blueshift velocity changes periodically and the period is $<$1 yr or $<$5 yrs\footnote{using different estimations of the black hole mass} using the equation of \cite{Eracleous97}, which are inconsistent with the variability behavior of the BEL profiles.
The inconsistency was supported by long-term variability studies of AGNs with similar profiles \cite{Gezari07} which found their variability behaviors are also not consistent with orbital motion.

In addition to the disk models, there are other models such as binary broad line region and emission from a bipolar outflow \citep[][for a review]{Eracleous03}.
The binary broad line region model is also unlikely for SBS 1423+511 due to the disagreement between predicted and observed variability behaviours of the BEL profile.
Although there are observational evidences against the bipolar outflow model in other AGNs with double-peaked BELs (see section 6.4.2 in Eracleous \& Harpten 2003) which means that the model is not universal, this model is possible for SBS 1421+511 considering that the quasar has powerful radio jet and is accreting at a high rate, both of which are indicators of strong outflow.

According to these discussions, the model that SBS 1421+511 hosts a super-massive black hole which is recoiling at a radial velocity of $\sim$1400 \kmsb and is accompanied by a circular disk which emits double-peaked BELs can explain all the observed properties.
And at the same time other possible models exist such as emissions from a circular disk with spiral emissivity perturbations and emissions from a bipolar outflow.
Follow-up spectral monitoring of SBS 1421+511 is needed to set a better constraint on the velocity change rate.
Ultraviolet spectra are also needed to obtain the profiles of high ionized BELs and reveal possible broad absorption lines to examine the bipolar outflow model.
}

\subsection{The Seyfert-like Component of Feature N: Obscured AGN in the Companion or Gas Illuminated by Quasar?}
\label{4.3}

{ The Feature N consists of a Seyfert-like component in the center of the companion galaxy and a HII region in the outer region.
If we had not found Feature S, we were likely to make the conclusion that the Seyfert-like component of Feature N comes from an obscured AGN in the companion and hence SBS 1421+511 is a dual AGN system.
Feature S is similar to the Seyfert-like component in several aspects including size, velocity difference, emission line fluxes and flux ratios, implying they may have similar nature.
Furthermore, the high flux ratio of [OIII]/H$\beta$ of Feature S indicates that the quasar radiation is strong enough to illuminate the gas at such a distance, and thus the Seyfert-like component may also be illuminated by the quasar.
In addition, there are two facts which are unfavourable for an AGN in the companion.}
One is that powerful AGNs are rare, for example the fraction of galaxies with a [OIII] luminosity as high as $>$ $1.4\times10^8L_\odot$ is less than 5\% even in the most massive galaxies \citep{Kauffmann03}.
The other is that the electron density of { the Seyfert-like component} is $<$100 cm$^{-3}$, lower than the typical value of NLR which is several $10^2$ cm$^{-3}$.
{ \cite{Xu07} investigated a sample of Seyfert 1 galaxies and found that nearly none of Seyferts with BEL FWHM(H$\beta$) $>$ 2000 \kmsb has NLR with such low density, and the fraction in narrow line Seyfert 1 galaxies is only $\sim$20\%.
Though we can not ruled out the possibility of a dual AGN in SBS 1421+511, we should consider other interpretations for the Seyfert-like component.}

One possible interpretation is that { the Seyfert-like component of Feature N} also comes from an EELR and just coincides with the companion galaxy.
It is reasonable that Feature S has an corresponding feature in the opposite direction considering that all possible origins of the EELRs, including tidal features, large scale radio jets and large scale outflows may have bipolar structure.
The existence of EELRs with bipolar structure in other quasars, e.g. SDSS J1356+1026 \citep{Greene12}, gives weight to this speculation.
In addition, the rareness of powerful AGN and the low electron density can also be naturally explained.

Another possible interpretation is that { the Seyfert-like component of Feature N} comes from the ISM in the companion galaxy illuminated by the quasar \citep{Filippenko85}.
The companion galaxy located in the ionization { cone} of a quasar would be ionized and look like a giant HII region for a distant observer if the quasar radiation is powerful enough.
In order to test whether the quasar in SBS 1421+511 is capable of generating the observed line ratios and [OIII] luminosity, we ran a simulation using the photo-ionization code \textbf{Cloudy} \citep{Ferland98}.
{ We first attempted an oversimplified single-component model} with input conditions as follows:
1. The shape of the quasar spectrum was obtained by interpolating the SED of 3C206 and normalized to the bolometric luminosity of the quasar.
2. The distance $d$ was estimated using the projected distance 12.5 kpc and assuming the unknown inclination angle to be the representative value $i=45^\circ$.
3. The density was assumed to be constant, and the value was set as the mean value of the ISM for a disk galaxy of $n_{\rm H}\sim1$ cm$^{-3}$.
4. The metallicity was assumed to be the solar one.
The simulation shows that the quasar radiation can produce a kpc-thick HII region.
The predicted line ratio of [OIII] $\lambda$5007/H$\beta$ $\sim$ 10.
{ The [OIII] emission of AperN comes from a region in a rectangular aperture which is set by 0.75$\arcsec$ slit and 2$\arcsec$ width AperN, which has a cross-sectional area of $A=26kpc^2$.}
Assuming the covering factor of the emitting gas, which depends on the orientation of the disk, is $cf\sim\frac{A}{4\pi d^2}$, the model predicts that [OIII] luminosity is $\sim2\times10^8\ L_\odot$.
{ As both the [OIII] luminosity and the flux ratio of [OIII]/H$\beta$ are high enough for the observed value, this oversimplified single-component model shows that the quasar radiation is strong enough to illuminated the ISM of the companion.}
However, the model predicted that the line ratio of [NII] $\lambda$6584/H$\alpha$ = 0.2, which is significantly lower than the observed value.
We attempted to vary the distance, density and metallicity in reasonable ranges, but failed to reproduce the data, for example a density of $\sim10^2$ cm$^{-3}$ and a super-solar metallicity could generate a high enough [NII]/H$\alpha$ ratio but the [OIII]/H$\beta$ ratio is too low.
The inconsistency { is common when one attempts to reproduce the spectrum of EELRs using photo-ionization models}, and is removed by considering at least two kinds of clouds with significantly different physical properties \citep[e.g.][]{Viegas92}.
For example, though the parameters are not well limited, a double-component model consisting of a low density component with a density of $\sim1$ cm$^{-3}$ and a solar or sub-solar metallicity, and a high density component with a density of $\sim10^2$ cm$^{-3}$ and a super-solar metallicity can produce enough [NII] and [OIII] at the same time.
And the two components could be explained as the warm ionized medium in the disk and high density medium in the galaxy center respectively for a typical galaxy.
Thus the interpretation that the Seyfert-like component comes from gas in the companion illuminated by the quasar is supported by the simulation.
This interpretation does not require that the companion galaxy is active and the low electron density of { the Seyfert-like component} is more natural.
However the 70 \kmsb velocity shift between [OIII] emission line and the starlight absorption lines of the companion is difficult to be explained.
{ By comparison, the shift can be easily explained if the Seyfert-like component} comes from an AGN because the shift is common among AGNs or { if it comes from an EELR} because the companion and the EELR are not necessarily associated.

{ In the survey of dual AGNs, separated [OIII] emission corresponding to the companion of a quasar is frequently used as indicator of an obscured AGN in the companion.
It is important to make clear whether SBS 1421+511 is a dual AGN because the conclusion may influence the strategies of survey of dual AGNs.}
Because [OIII] illuminated by an AGN of the companion galaxy would concentrate on the center, while that illuminated by quasar would be more extended, [OIII] distribution map could be used to determine whether the companion is active.
However, we can not make an unambiguous distinction based on ESI data taken under 1.4$\arcsec$ ($\sim$6 kpc) seeing condition.
Thus future narrow band imaging or IFS observations with high spatial resolution are required to map the [OIII] distribution, hence make a final conclusion.

\subsection{Possible Origins of the EELRs}
\label{4.4}

We first ruled out the possibility that the EELRs are generated by large scale radio structures because the radio source is compact and shows no large scale features.
Since we have seen a kpc-scale outflow from the blue wing of quasar NELs, a natural question is that whether it is related with the EELRs.
We checked whether the kpc-scale outflow can provided enough material to form a structure like Feature S.
The total mass of Feature S is estimated to be $M\sim3\times10^6M_\odot$ using the electron density from [SII] doublet and H$\beta$ luminosity.
However, the situation may be more complicated because a single-component model either can not reproduce the observed line ratios for Feature S as what happened for the Seyfert-like component of Feature N: a similar \textbf{Cloudy} simulation shows that a cloud with a density of $n_e=120$ cm$^{-3}$ would not reach a high enough ionization state to produce as much [OIII] relative to H$\beta$ as observed, and again lower densities which can produce enough [OIII] fail to produce as much [NII] and [SII] as observed.
Thus a double-component model is also needed for Feature S and the electron density from [SII] doublet only applies for the high density component since [SII] emission mainly comes from it.
If adopting this model, the total mass would be higher than what we estimated previously because the low density component would be much more massive assuming that the two components contribute comparable H$\beta$.
For example, \cite{Stockton02} showed that the EELR of quasar 4C 37.43 is better modeled as a two-phase medium, consisting of a matter-bounded diffuse component ($n\sim2$ cm$^{-3}$) and an embedded small, dense component ($n\sim500$ cm$^{-3}$), and the fractions of the flux of H$\beta$ from the two components are $\sim$1/3 and $\sim$2/3, respectively.
If this is the case for SBS 1421+511, the mass of low density component would be about two orders of magnitude higher than that of the high density component, and the total mass would be $\sim10^8M_\odot$.
{ The dynamic time scale of an outflow capable of producing an emission line feature like Feature S is $t_d=10^8$ yr, which is estimated using the size and mean radial velocity of Feature S.}
Considering the mass rate of the kpc-scale outflow and the dynamic time scale, the blue wing outflow can provide enough material to form an EELR feature like Feature S.
However { the model has a disadvantage}: a cloud with a density of $\sim10^2$ cm$^{-3}$ and a temperature of $\sim10^4$ K in circum-galactic environment is unstable and would expand due to thermal pressure.
For example, the pressure is $\sim10^2$ higher than that of surrounding hot ionized medium, which has a typical density of $<10^{-2}$ cm$^{-3}$ and a temperature of $\sim10^6$ K.
The gas is also unlikely to be gravitational condensed because the minimum mass for a cloud to reach a equilibrium between gravity and thermal pressure is $\sim10^7M_\odot$, which is unlikely to be true.
Using the Str\"{o}mgren radius of $\sim0.1$ pc estimated from ionization parameter and density of Feature S, and the sound speed $v\sim\sqrt{kT/m_p}\sim10$ \kms, the life time of a high density cloud is estimated to be $\sim10^4$ yr, far shorter than the dynamic time scale of an outflow which is expected to form Feature S.
Thus there must be a mechanism to continuously regenerate the observed high density clouds if Feature S indeed comes from the outflow.

Considering that SBS 1421+511 is likely to be a merging system, an alternative possible origin is that the EELRs are tidal tails which are illuminated by quasar.
Apart from the suggestion by the close projected distance { and little velocity difference}, there are also two pieces of evidence of strong interactions between the two galaxies.
One is that the residual of SDSS 2-D spatial decomposition (Figure 3(f)) shows a feature in northwest direction from the center of the companion galaxy.
The other is that the companion galaxy shows an off-center HII region, which is rare among isolated galaxies and can be naturally interpreted as a tidal tail { of the companion galaxy}.
{ One can imagine the natures of the two emission line features of SBS 1421+511 according to nearby mergers: Feature N comes from the bridges between the two galaxies and the tidal tails of the quasar companion, and Feature S comes from the tidal tails of the quasar host.
Thus the two features can be understand in a unified scenario.}

Since EELRs associated with tidal features generally show correspondence between the continuum and emission-line morphologies while those ejected by outflow unnecessarily do, the distribution maps of emission line and continuum may help distinguish the two possible origins, tidal or outflow.
However we can not make a detailed analysis based on Keck data taken under 1.4$\arcsec$ seeing condition.
New observations with higher spatial resolution are required, which would map the structure and dynamic of the system, and thus help to clarify the origin of the EELR in SBS 1421+511.

\section{Summary and Future Observations}

{ We selected SBS 1421+511 from SDSS DR7 quasar catalog as a recoiling super-massive black hole candidate for that the BELs show blue-skewed profile with a peak velocity of $\sim$3300 \kmsb blueshifted relative to NELs and that there is offset between the quasar and the nearby extended source.}
We presented results of Keck ESI long slit spectrographic observation of SBS 1421+511.
By applying spectral and spatial decomposition technics, we identified an host galaxy at the quasar position and a companion galaxy 3$\arcsec$ north to the quasar with velocity difference of 120 \kmsb relative to the host galaxy.
Thus the off-nuclear AGN character is not due to a recoiling black hole, but due to a close companion of the quasar.
{ We fitted the profile of H$\alpha$ BEL using disk models, which are the most successful model for AGNs with similar profiles.
A circular disk model can well fit the data only if applying an ad hoc blueshift of $\sim$1400 \kms, and the elliptical disk model is ruled out.
We detected no change of the peak velocity of the profile among MDM, SDSS and Keck observations, setting a 3$\sigma$ upper limit of the velocity change rate of 14 \kms per year, and thus the period of the velocity change is $>$200 yrs.
The non-detection of velocity change is inconsistent with the three models: emission from a circular disk with an orbiting bright spot, emission from a circular disk which orbiting around an unseen companion and emission from binary broad line regions.
Remained are three models: emission from a circular disk which is accompanied with a super-massive black hole recoiling at a velocity of $\sim$1400 \kms, emission from a circular disk with spiral emissivity perturbations and emission from a bipolar outflow, and the first is the most likely.
Thus SBS 1421+511 may hold a supermassive recoiling black hole.}

The companion galaxy shows Seyfert-like line ratios and strong [OIII] emission with a luminosity of $>1.4\times10^8L_\odot$, suggesting that it may also be an AGN and hence SBS 1421+511 is a dual AGN system.
Intriguingly, an EELR is detected in the opposite direction of the companion southern to the quasar and thus the companion galaxy may only appear to be active.
There are two possible interpretations.
One is that the EELR has a bipolar structure and the northern half just coincide with the companion galaxy.
The other is that the gas in the companion is illuminated by the quasar other than its own AGN, which is supported by a simulation with photo-ionization model.

{ The unsolved mysteries in SBS 1421+511 need follow-up observations.
Future optical spectral monitoring and ultraviolet spectrometry are needed to make a better understand on the nature of the BEL of SBS 1421+511 and test whether it host a super-massive recoiling black hole.
Future narrow line imaging or IFS observations with sub-arcsec ($<$1 kpc) spatial resolutions are required to reveal possible merging features to determine whether the two galaxies are physically connected, and to map the [OIII] brightness distribution to tell whether the companion is active.}

\begin{acknowledgements}
The data presented herein were obtained at the W. M. Keck Observatory, which is operated as a scientific partnership among the California Institute of Technology, the University of California and the National Aeronautics and Space Administration.
The Observatory was made possible by the generous financial support of the W. M. Keck Foundation.
We are grateful to the anonymous referee for his/her helpful comments and for pointing out possibilities of the emission line features.
We also would like to thank J. X. Prochaska for helping us learn Keck observations and data reduction using his program, and thank G. Li for offering language assistant.
This work is supported by the NSFC grant (11421303, 11473025, NSF11033007), National Basic Research Program of China (973 Program, 2013CB834905), the SOC program (CHINARE2012-02-03), and Fundamental Research Funds for the Central Universities (WK 2030220010).

\end{acknowledgements}

\begin{table}[!t]\footnotesize
\caption{Parameters of the disk models for SBS 1421+511.}
\begin{threeparttable}
\begin{tabular}{ccccccccc}
\hline
\hline
model          &$\chi^2$/dof &$v_{bs}$ (km/s) &$\xi$ (R$_s$) &i (deg) &q &$\sigma$ (km/s) &$\phi_0$ (deg) &e\\
\hline
circular                &19644/844  &     &570--5900 &38$^\circ$ &2.0 &660  &&\\
elliptical              &6466.7/842 &     &290--5500 &41$^\circ$ &2.0 &1540 &11$^\circ$ &0.35\\
circular (blueshifted)   &1012.8/843 &1450 &150--1300 &17$^\circ$ &1.7 &1090 &&\\
elliptical (blueshifted) &1003.9/841 &1230 &180--1200 &21$^\circ$ &1.8 &790  &12$^\circ$ &0.18\\
\hline
\hline
\end{tabular}
\end{threeparttable}
\label{tab1}
\end{table}

\begin{table}[!t]\footnotesize
\caption{Line ratios of the extended emission line features}
\begin{threeparttable}
\begin{tabular}{cccc}
\hline
\hline
                      &[OIII]/H$\beta$ &[NII]/H$\alpha$ &[SII]/H$\alpha$\\
\hline
\multicolumn{4}{c}{off-center HII region}\\
\hline
AperHII\tnote{a}      &$2.62\pm0.14$   &$0.65\pm0.04$   &$0.32\pm0.02$   \\
AperHII\tnote{b}      &$1.21\pm0.17$   &$0.33\pm0.05$   &$0.20\pm0.03$   \\
HII-like set\tnote{c} &$0.39\pm0.06$   &$0.31\pm0.02$   &$0.15\pm0.02$   \\
\hline
\multicolumn{4}{c}{Seyfert-like component of Feature N}\\
AperN\tnote{a}        &$6.27\pm0.15$   &$1.06\pm0.03$   &$0.60\pm0.02$   \\
AGN-like set\tnote{c} &$7.04\pm0.23$   &$0.88\pm0.02$   &$0.62\pm0.02$   \\
\hline
\multicolumn{4}{c}{Feature S}\\
\hline
AperS\tnote{a}        &$5.98\pm0.22$   &$0.87\pm0.04$   &$0.62\pm0.03$   \\
\hline
\hline
\end{tabular}
\begin{tablenotes}
    \item [a] line fluxes are obtained by performing Gaussian fitting to the corresponding spectra.
    \item [b] line fluxes are calculated by integrating flux over velocity range of $-240<v<-80$ \kms.
    \item [c] line fluxes are obtained from the decomposition results of the two sets.
\end{tablenotes}
\end{threeparttable}
\label{tab2}
\end{table}

\begin{figure}
\centering{
 \includegraphics[scale=0.98]{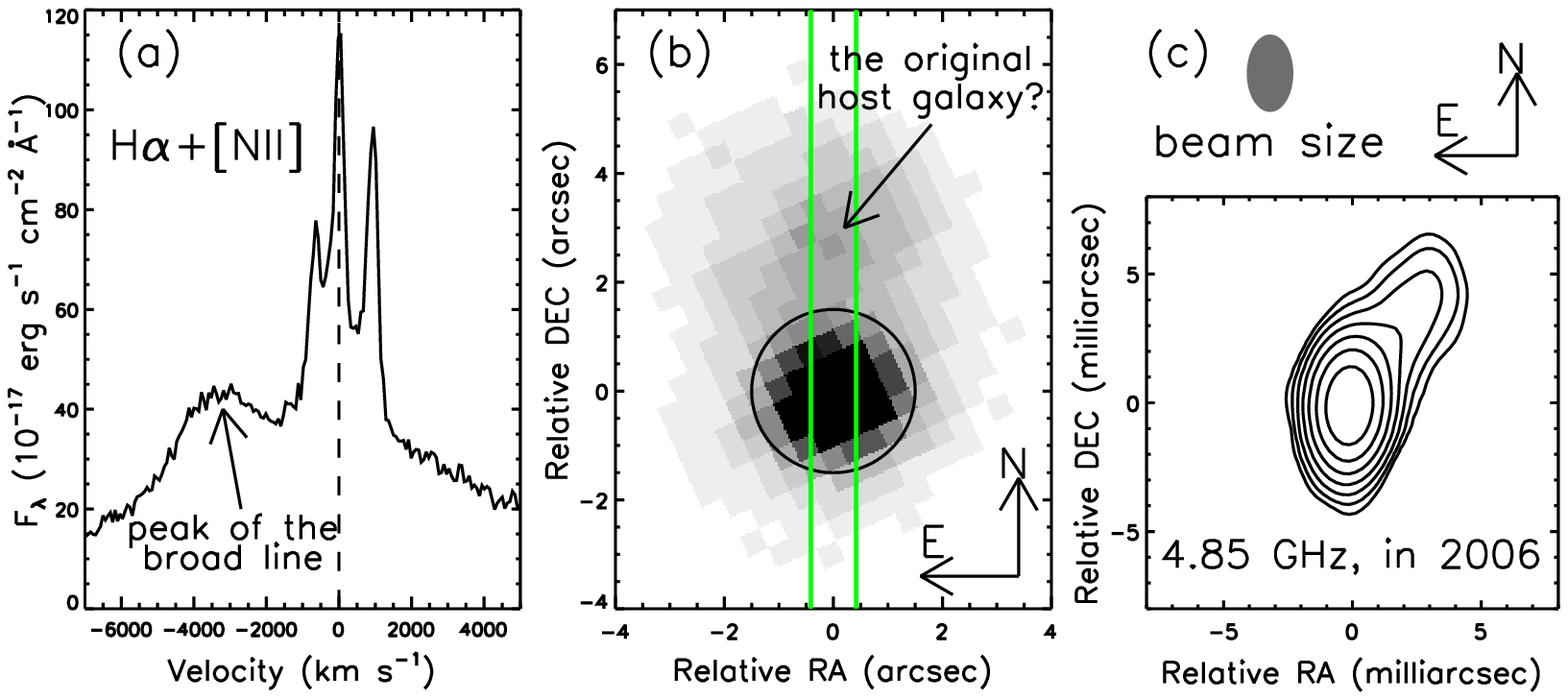}
 \caption{:{
  \textbf{(a)}: SDSS spectrum in H$\alpha$+[NII] $\lambda\lambda$6548, 6583 region.
  The wavelengths are converted to velocities according to H$\alpha$ NEL.
  \textbf{(b)}: SDSS five-band coadded image.
  The aperture for the SDSS spectrometry and the slit for the Keck observation are illustrated with a black circle and a pair of green lines, respectively.}
  \textbf{(c)}: VLBA brightness distribution map of the quasar at 4.85 GHz taken on 2006 August 10.
  Contours are plotted according to levels from $2^{-1}$ to $2^{-7}$ of the peak flux.
  }}
 \label{sdss}
\end{figure}

\begin{figure}
\centering{
 \includegraphics[scale=0.98]{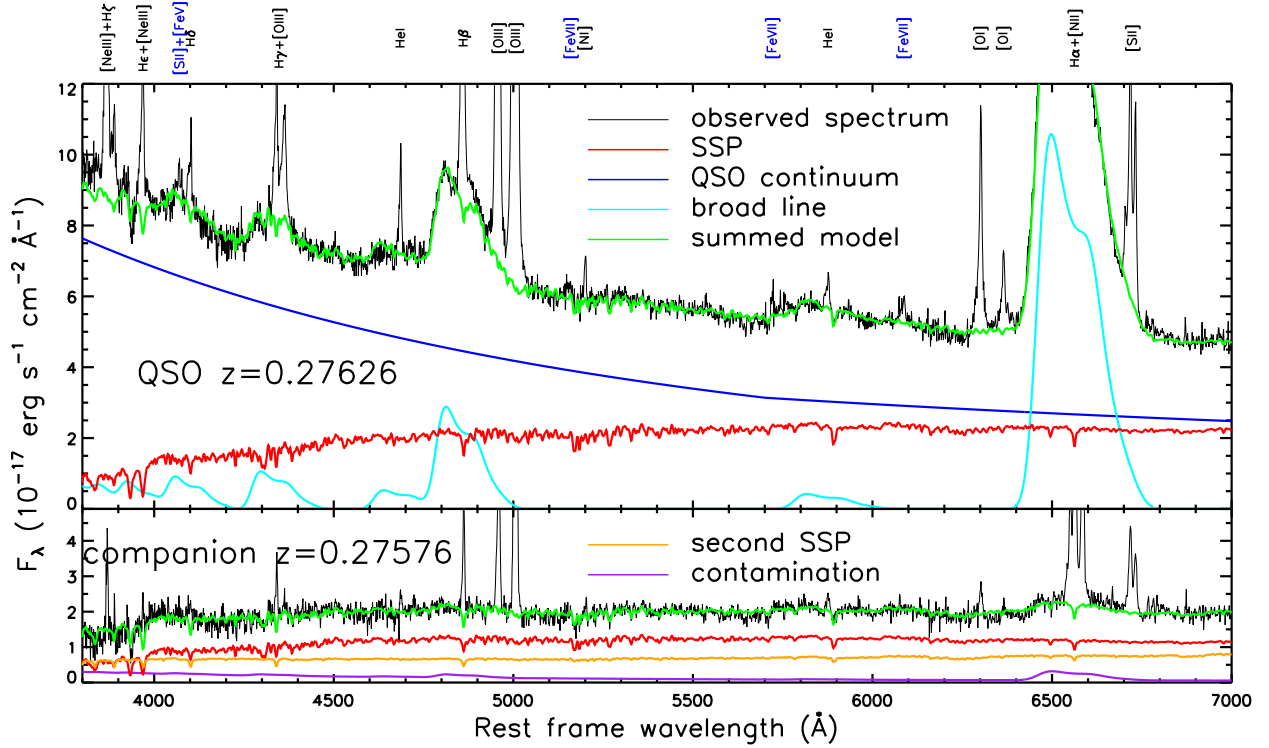}
 \caption{:
  The spectra of the quasar and the companion, along with the best-fitting continuum models.
  These are transferred to rest frame using stellar redshifts obtained from fitting and binned by 7 pixels for demonstration.
  For the quasar spectrum, the model consists of a broken powerlaw for the quasar continuum component (blue), a single-SSP for the stellar component (red) and a BEL component (cyan).
  For the companion galaxy, we fit the spectrum using a double-SSP for the stellar component (red and orange) and a quasar contamination component (purple).
  We label the detected NELs at the top using the line list from Vanden Berk et al. (2001), and the lines which are only detected in the quasar spectrum are marked in blue.
  We masked the regions around the NELs using velocity ranges $-1400<v<800$ \kmsb and $-600<v<400$ \kmsb for the quasar and the companion, respectively.
  }}
 \label{starlight}
\end{figure}

\begin{figure}
\centering{
 \includegraphics[scale=0.98]{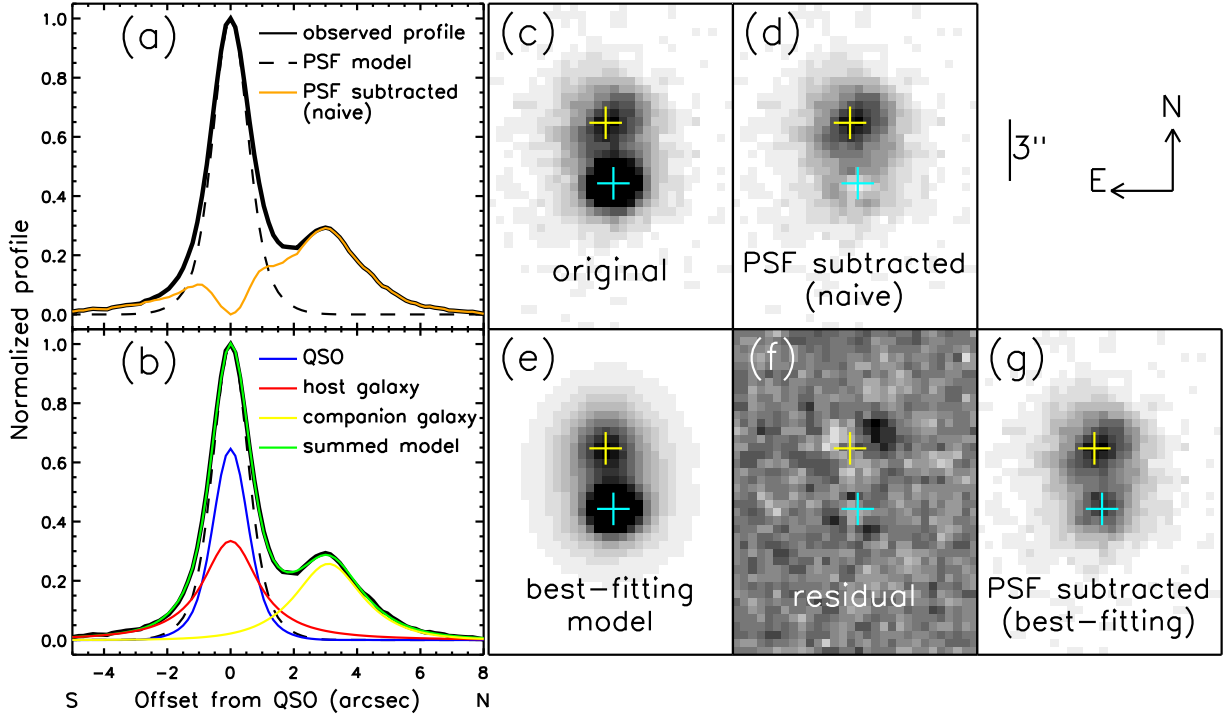}
 \caption{:
  \textbf{(a)}:
  Black line shows the spatial brightness profile of the continuum around rest-frame 7000 \AA, { which is normalized at the quasar position.
  To demonstrate the underneath stellar component, we subtracted the contribution of the quasar, which is obtained with PSF (dashed line), from the original profile.
  The orange line shows a residual by subtracting an quasar component so that the residual flux at quasar position is zero.
  This ``naive'' subtraction, which obviously overestimates the quasar flux, represents the minimum flux of the stellar component.}
  \textbf{(b)}:
  { the 1-D profile decomposition results.
  The model consists of a quasar component (blue), a S\'{e}rsic profile for the host galaxy (red) and another S\'{e}rsic profile for the companion galaxy (yellow).
  The summed model is shown in green.}
  \textbf{(c)--(g)}:
  { SDSS i-band image (c), the residual of a similar naive subtraction (d), and the 2-D profile decomposition results (e--g) using the same model as the 1-D decomposition.}
  The quasar position (cyan) and the center of the companion galaxy (yellow) obtained from decomposition are marked with plus.
  }}
 \label{spatial}
\end{figure}

\begin{figure}
\centering{
 \includegraphics[scale=0.98]{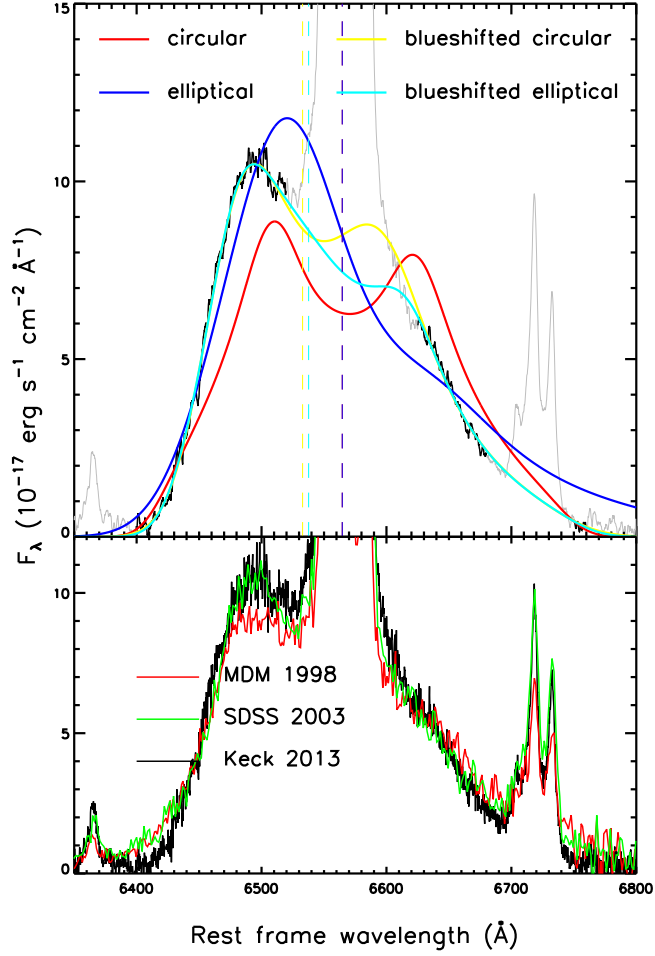}
 \caption{:
  { The upper panel:
  The emission line spectrum of the quasar in H$\alpha$ spectral region and the best-fitting disk models.
  We fit the H$\alpha$ BEL in three spectral regions (6400--6520 \AA, 6625--6680 \AA\ and 6750--6760 \AA) which are not affected by NELs.
  The spectrum in the three regions are plotted in black and that out of the regions are plotted in grey.
  We overplotted the results of four disk models: circular disk with (yellow) or without (red) a blueshift, and elliptical disk with (cyan) or without (blue) a blueshift.
  The lower panel:
  The H$\alpha$ profiles from MDM, SDSS and Keck observations, which are normalized in spectral region of 6620--6680 \AA.}
  }}
\label{dpfit}
\end{figure}

\begin{figure}
\centering{
 \includegraphics[scale=0.98]{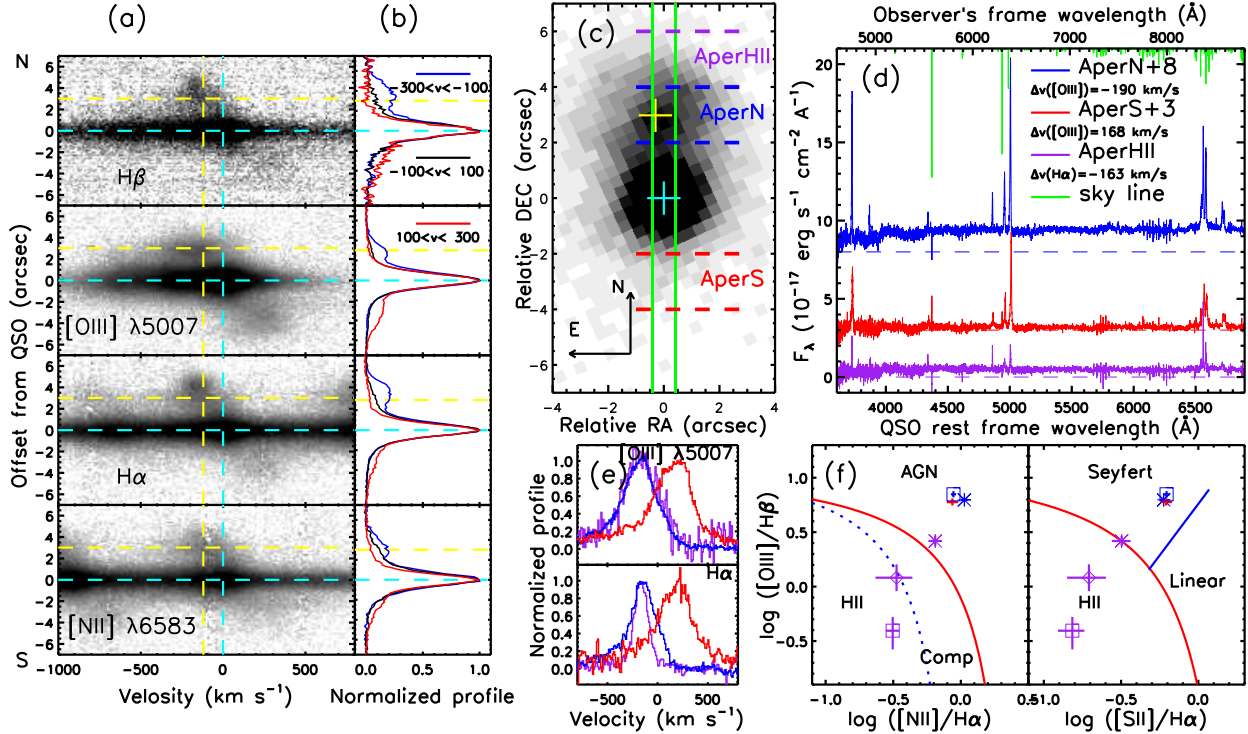}
 \caption{:
  \textbf{(a)}: 2-D emission line spectral image in four regions in logarithmic flux units, revealing features in both the two directions, { after subtracting the continuum from the original spectral image.
  For each spectral region, the continuum was obtained by performing local second-order polynomial fit in nearby regions free of emission lines.}
  The spectral direction is horizontal, while the spatial direction is vertical (north is up).
  The horizontal dashed lines represent the quasar position (cyan) and companion galaxy center (yellow), and the vertical lines represent the velocities of starlight of the two galaxies.
  \textbf{(b)}: Normalized spatial brightness profiles within three velocity ranges.
  Ranges $-300<v<-100$ \kmsb (blue) and $100<v<300$ \kmsb (red) correspond to Feature N and Feature S, respectively.
  Profiles within velocity range $-100<v<100$ \kmsb (black) are also plotted for comparison.
  \textbf{(c)}: Same with Figure 1(b), illustrating the dividing lines (dashed) of the three apertures: AperN (blue), AperHII (purple) and AperS (red).
  \textbf{(d)}: 1-D spectra within the three apertures from the original spectral image, plotted in corresponding colors with labeled velocity differences.
  \textbf{(e)}: The velocity (relative to the quasar) profiles of [OIII] $\lambda$5007 and H$\alpha$ lines for the three spectra.
  The contributions of other NELs (e.g. [NII] $\lambda\lambda$6548,6583 doublets when plotting H$\alpha$), the continua, and the contaminations are subtracted using the best-fitting models.
  \textbf{(f)}: Flux ratios of emission lines plotted over the diagnostic diagram of \cite{Kewley06}.
  The different symbols are described in the text.
  The error bars correspond to 95.3\% confidence.
  }}
 \label{eelr}
\end{figure}

\begin{figure}
\centering{
 \includegraphics[scale=0.98]{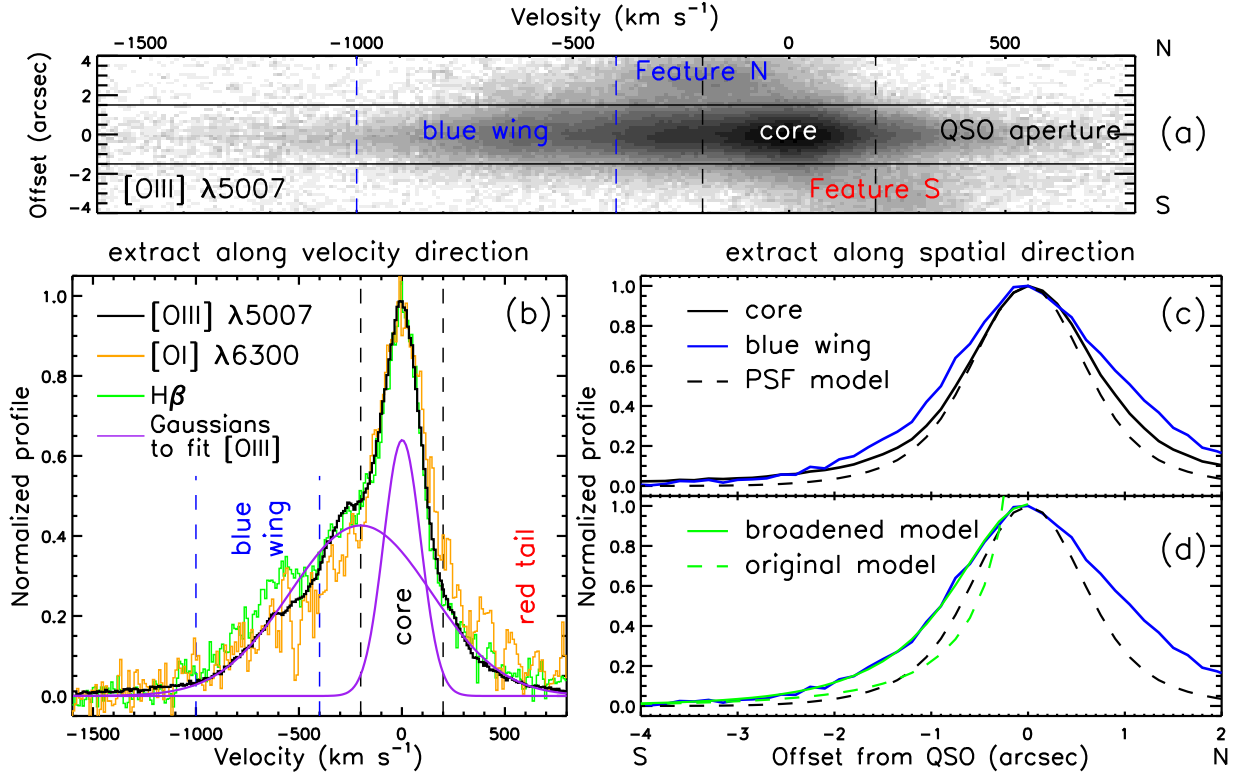}
 \caption{:{
  \textbf{(a)}: 2-D spectral image in [OIII] $\lambda$5007 spectral region from which continua and BELs were subtracted using best-fitting models.
  Note that the north half of the blue wing is influenced by Feature N.
  \textbf{(b)}: The normalized velocity profile of [OIII] within quasar aperture (black), along with the double Gaussian model (purple).
  We defined two velocity ranges corresponding to the core component and the blue wing component, respectively, which are illustrated in dashed line.
  In addition, the profiles of H$\beta$ and [OI] $\lambda$6300 are plotted for comparison.
  \textbf{(c)}: Normalized spatial brightness profile of the core component (black) and the blue wing component (blue) for [OIII].
  The PSF is shown in dashed line for comparison.
  \textbf{(d)}: Fitting results for the south half of the spatial profile of the blue wing component.
  The best-fitting S\'{e}rsic model which has been broadened by convolving the PSF is plotted in green solid line, and the original model is plotted in green dashed lines.
  }}}
\label{outflow}
\end{figure}

\end{document}